\documentclass[aps,pre,twocolumn,amsfonts,amssymb,amsmath,showpacs,floatfix, superscriptaddress]{revtex4-1} 
\usepackage{color}
\usepackage{graphicx}

\usepackage{graphicx}
\newcommand{\CC}{\ensuremath{\mathcal{C}}}
\newcommand{\de}{\delta} % for functional derivative
\newcommand{\XX}{\mathbf{X}}
\newcommand{\RR}{\mathbf{R}}
\newcommand{\dd}{\partial}
\newcommand{\uu}{\mathbf{u}}
\newcommand{\s}{\sigma}
\newcommand{\nn}{\nonumber}
\newcommand{\bP}{\mathbf{P}}
\newcommand{\DD}{\mathcal{D}}
\newcommand{\eps}{\epsilon}

\begin{document}

\title{Variational principle for non-linear wave propagation in dissipative systems
%\\ in inhomogeneously anisotropic media
}
\author{Hans Dierckx}
\affiliation{Department of Mathematical Physics and Astronomy, Ghent
  University, 9000 Ghent, Belgium}

\author{Henri Verschelde}
\affiliation{Department of Mathematical Physics and Astronomy, Ghent
  University, 9000 Ghent, Belgium}

\date{\today}

\begin{abstract}
The dynamics of many natural systems is dominated by non-linear waves propagating through the medium. We show that the dynamics of non-linear wave fronts with positive surface tension can be formulated as a gradient system. The variational potential is simply given by a linear combination of the occupied volume and surface area of the wave front, and changes monotonically in time. Finally, we demonstrate that vortex filaments can be written as a gradient system only if their binormal velocity component vanishes, which occurs in chemical systems with equal diffusion of reactants. 
\end{abstract}

\maketitle

\textit{Introduction. } 
Propagating non-linear waves are found in many extended systems, which can be of oscillating, bistable or excitable nature. Chemical examples include combustion \cite{Zeldovich:1938}, catalytic oxidation \cite{Nettesheim:1993}, and spiral patterns in oscillating chemical reactions \cite{Zhabotinsky:1973}. In physics, dynamical phase transition boundaries are also known as domain walls \cite{Weinberg:1995}. In biological context, non-linear waves often fulfil an intra- or extracellular signalling role. They have been observed both on the micrometer scale, e.g. as calcium waves \cite{Lechleiter:1991}, and on a macroscopic scale for action propagation in neural \cite{Hodgkin:1952} and cardiac \cite{Klabunde:2004} tissue, where they govern cognitive processes and coordinated contraction of the heart.  

All above-mentioned systems share the emergent phenomenon of non-linear wave propagation. In contrast to classical electromagnetism, non-linear waves have a fixed amplitude and do not linearly superimpose when colliding; instead they may annihilate or reflect \cite{Aslanidi:1997}. As a consequence, most non-linear waves (except solitons) lack the classical invariants of motion such as energy and momentum, which conservative systems do possess. This lack is no contradiction with the laws of thermodynamics, since generally energy is consumed in the process of wave propagation in the form of heat, which is why these waves are also referred to as dissipative structures \cite{Prigogine:1984}. Due to the lack of invariants of motion, the analytical and physical description of dissipative systems remains a challenge. 

Some progress has been made in rewriting the partial differential equations (PDEs) for specific dissipative systems as a so-called gradient system \cite{Hirsch:2004}. Here, we are interested to write a general system as a continuous gradient system, which is a set of PDEs that can be written in the form \cite{Mornev:1998}:
\begin{align}
  \dot \XX(s,t) = - \frac{\de F}{\de \XX}, \label{gradsys}
\end{align}
for a functional $F = \int_{\Omega} f(\XX, \dd_s \XX,  s) ds$. Here $\XX$ is a state vector and $s$ represents a set of parameters labelling a $N-$dimensional spatial domain $\Omega$. The dot denotes the partial derivative with respect to time and the notation $\delta . / \delta .$ refers to the functional derivative \cite{Frankel:1997}.

Note that \eqref{gradsys} immediately implies that
\begin{align}
  \frac{d F}{d t} = \int \frac{\de F}{\de \XX} \cdot \dot\XX ds = - \int \left(\frac{\de F}{\de \XX} \right)^2 ds  \leq 0.  \label{decrease}
\end{align}
Hence the potential $F$ decreases in time and therefore is a Lyapunov function of the system. Thus, if a critical point is an isolated minimum of $F$, it will be an asymptotically stable equilibrium point \cite{Hirsch:2004}. In a sense, the dynamics of gradient systems is relatively easy to understand, since finding the potential function $F$ at once reveals the long-term attractors of the dynamical system. However, finding the potential for a given system remains much of an art, and it is unlikely that a single method will be successful for any PDE that describes a dissipative system.  

We illustrate the difficulty of finding a gradient system form on the important class of reaction-diffusion systems. These dissipative systems are modelled by a set of parabolic PDEs, which are in one spatial dimension of the form:
\begin{align}
 \dot \uu(x,t) = \bP \dd_x^2 \uu(x,t) + \RR(\uu(x,t)).  \label{rde}
\end{align}
A necessary condition under which a system can be recast as a gradient system is found by taking the functional derivative of both sides in Eq. \eqref{gradsys}. Letting the state variables $u^j$ act as the components of $\mathbf{X}$ in \eqref{gradsys}, we find
\begin{align}
  \frac{ \de \dot{u}^i}{ \de u^j} =   \frac{ \de \dot{u}^j}{ \de u^i}.  \label{cond}
\end{align}
This compatibility condition is the functional analogue of the well-known theorem in vector calculus that $\vec{X} = \vec{\nabla} F \Leftrightarrow \vec{\nabla} \times \vec{X} = \vec{0}$ in a simply connected domain. For a reaction-diffusion system, the compatibility condition \eqref{cond} requires symmetry of the Jacobian of the reaction term. Thus, only for selected reaction terms $\RR(\uu)$, one may rewrite \eqref{rde} as a gradient system, e.g. for the diffusion equation $(\RR=\mathbf{0})$ or when there is only one state variable \cite{Mornev:1998}. With small amendments to the gradient system structure, the FitzHugh-Nagumo equations have been cast to this form \cite{Mornev:1992} and extensions with skew-symmetry have been considered in \cite{Yanagida:2002}.  However, condition \eqref{cond} implies that trying to rewrite every reaction-diffusion system as a pure gradient system is an impossible quest. 

In this Letter, we take a different approach by formulating not the underlying system but the emergent wave front dynamics as a gradient system. Therefore, our method will work for any dissipative system that supports traveling waves with uniquely selected wave speed and whose wave fronts end orthogonally to the medium boundaries. 

The dynamical law which we will formulate as a gradient system is the velocity-curvature relation for non-linear waves. In homogeneous media, the wave speed $c$, measured along the wave front normal $\vec{n}$, is found to depend on the wave front's extrinsic curvature $K$. When the wave front thickness is small compared to the front's local radius of curvature, the velocity-curvature relation can be linearized to 
\begin{align}
 c = c_0 - \gamma K  \label{cK1}
\end{align}
with plane wave speed $c_0$ and surface tension coefficient $\gamma$. For reaction-diffusion systems, the relation \eqref{cK1} can be derived from the underlying system equations \cite{Keener:1986, Dierckx:2011}. By convention, the speed $c_0$ is positive; it may vanish in certain dissipative structures \cite{Panfilov:1995b}. Further, stability analysis shows that only when the surface tension $\gamma$ is strictly positive, the non-linear waves are stably propagating and we will work in this regime. In this regime, small protrusions of the wave front will be dampened, such that after a short while, the front's curvature will be small, and it can be well described by Eq. \eqref{cK1}. 
 
\textit{Theorem. } Our main result in this work is that the geometric law of motion \eqref{cK1} for non-linear waves can be written as the continuous gradient system \eqref{gradsys} with ever-decreasing potential 
\begin{align}
  F(t) = - c_0 V(t) + \gamma S(t). \label{F}
\end{align}
where $V(t)$ is the volume where the traveling wave has passed and $S(t)$ is the wave front's total surface area. Our result holds for wave fronts in any number of spatial dimensions $N \geq 2$. In what follows, we will use terminology of surface and volume as if working in $N=3$ dimensions. 

\textit{Proof. } 
Our proof involves the calculus of variations and goes as follows. 
To describe the emergent dynamics, we first formalize the definition of a wave front. In a spatial domain that locally looks like $\mathbb{R}^N$, we identify a scalar quantity $u$ of the system (e.g. temperature or transmembrane potential) and fix the wave front at time $t$ as the hypersurface
\begin{align}
 \Omega_F(t) = \{ \vec{x}\  |\  u(\vec{x},t)=u_c \textrm{\ and\ } \dd_t u( \vec{x} , t ) >0  \}. 
\end{align}
The evolution of the wave front $\Omega_F$ can be described by coordinate functions $x^a = X^a(\s^A, t)$, where $a \in \{1,2,...,N \}$ and $A \in \{1,2,...,N-1\}$. Henceforth we will use the summation convention for repeated indices. Changing from Cartesian coordinates $x^a$ to curvilinear coordinates $\s^A$ induces a metric 
\begin{align}
 h_{AB} &= \dd_A x^a g_{ab} \dd_B x^b 
\end{align}
on the wave front with determinant $h$ and matrix inverse $h^{AB}$. In isotropic media, $g_{ab}$ is the identity matrix, while in systems with anisotropic diffusion it can be taken equal to the inverse diffusion tensor to capture spatially varying anisotropy \cite{Verschelde:2007, Young:2010}. The coordinate change allows writing the linear velocity-curvature relation \eqref{cK1} as:
\begin{align} \label{cK2}
    \dot X^a &= c_0 n^a + \gamma \DD^2 X^a,
\end{align}
with unit normal to the front 
\begin{align}
n_g &= \frac{\eps^{AB...F} \eps_{ab...g}}{\sqrt{h} (N-1)!} \dd_A X^a \dd_B X^b ... \dd_F X^f
\end{align}
where the $\eps^{...}$ are the fully antisymmetric Levi-Civita symbols. Furthermore, in Eq. \eqref{cK2}, $\DD^2 = (\sqrt{h})^{-1}\dd_A(\sqrt{h} h^{AB} \dd_B) $ is a $N-1$ dimensional covariant Laplacian with respect to the tangential coordinates $\s^A$. 

\begin{figure}
 \includegraphics[width=0.45 \textwidth]{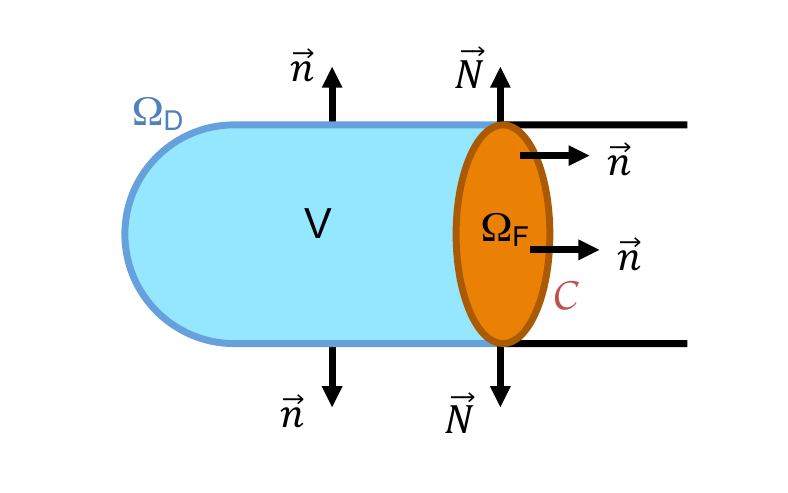}
\caption{ (Color online) Geometry considered for the variational principle. The wave front $\Omega_F$ (red) with unit normal vector $\vec{n}$ has passed in all points of the volume $V$. The interface of $V$ with the domain boundary is a surface $\Omega_D$ with outward normal vector $\vec{n}$. Only at the edge $\CC$ of the wave front, the vector field $\vec{n}$ is discontinuous; there we define $\vec{N}$ as the outward normal vector to the medium. \label{fig:geom} }
\end{figure}

To prove our theorem \eqref{gradsys}, \eqref{F}, we have to show that
\begin{align}
  \frac{\de V}{\de \vec{X}} &= \vec{n}, &
  \frac{\de S}{\de \vec{X}} &= K\vec{n} \label{toproof}
\end{align}
under appropriate boundary conditions. When the wave front touches the domain boundary, the surface bordering the volume where the wave has passed can be divided into the wave front surface $\Omega_F$ and the surface $\Omega_D$ where the occupied volume touches the domain boundary. This situation together with the outer normal vectors is sketched in Fig. \ref{fig:geom}. The $N-2$ dimensional border between $\Omega_F$ and $\Omega_D$ is $\CC$, the wave front boundary. In this work, we will not consider broken wave fronts (scroll waves), and assume that wave fronts only end on the domain boundary. In this case, the volume where the wave has passed has a border $\Omega = \Omega_F \cup \Omega_D \cup \CC$ that is a closed surface. We will assume that the wave front intersects the boundary orthogonally, such that we can impose:
 \begin{eqnarray}
\forall \vec{x} \in \Omega_F&: & \de \vec{X}\  ||\  \vec{n}, \nn 
\\
\forall \vec{x} \in \Omega_D&: & \de \vec{X}\ \cdot \vec{n} = 0, \label{BC} \\
\forall \vec{x} \in \CC&: & \de \vec{X} \textrm{ is \ continuous across } \CC. \nn 
\end{eqnarray} 
The first condition is the transverse gauge, in which we convene without loss of generality that points of the wave front move orthogonally to it. The second condition ensures that the wave front does not propagate across the domain boundary, but allows reparameterization of the interface. The third condition is compatible with the first two only when the wave front ends orthogonally on the domain boundary. This situation automatically follows if the physical boundary is impermeable to the reactants of the underlying system and if the spatial variations in the domain boundary are of length scales bigger than the wave front thickness. 

In a first step, we construct the volume occupied by the wave front. Take the origin of the Cartesian coordinate system in the region where the front has passed. The volume of the cone with the apex at the origin and an elementary surface area of the wave front as the base is $dV = (1/N) \vec{X} \cdot \vec{n} dS$ with $dS = \sqrt{h} d^{N-1}\s$, whence 
\begin{align}
  V = \oint_\Omega dS \frac{\eps^{AB...F}}{N! \sqrt{h}} \eps_{ab..g} \dd_A X^a \dd_B X^b ... \dd_F X^f X^g. \label{actionV}
\end{align}
We have used the symbol $\oint$ to emphasize that the integration runs over the closed surface $\Omega$. Under small variations $\de \vec{X}(\s^A)$ in the direction of the wave front normal only, the resulting change of  volume is the functional differential
\begin{align}
\de V =& \frac{\eps^{AB..F} \eps_{ab..g} }{N!} \oint_\Omega d^{N-1} \s [ \dd_A \de X^a .... \dd_F X^f \de X^g \nn \\ 
&+ .... + \dd_A X^a .... \dd_F \de X^f X^g + \dd_A X^a ... \dd_F X^f \de X^g ]  \nn \\
=& \frac{\eps^{AB..F} \eps_{ab..g} }{N!} \oint_\Omega d^{N-1} \s [ - \de X^a .... \dd_F X^f \de \dd_A X^g \nn \\ &-  .... - \dd_A X^a ....  \de X^f \dd_F X^g + \dd_A X^a ... \dd_F X^f \de X^g ] \nn \\
= &  \frac{\eps^{AB..F} \eps_{ab..g} }{(N-1)!} \oint_\Omega \frac{ d^{N-1}\s}{\sqrt{h}}  \dd_A X^a ... \dd_F X^f \de X^g \nn \\
=& \oint_\Omega dS \ \vec{n} \cdot \de \vec{X}.  \label{dV}
\end{align}
In the first equality, we performed integration by parts and the boundary terms always vanish due to integration over a closed surface. For the second equality, we renamed $a \leftrightarrow g$ in the first term, followed by a transposition in the Levi-Civita symbol, such that the alphabetic ordering of indices is restored. Doing this for the (N-1) first terms yields the desired result. The last line of Eq. \eqref{dV} implies that $\de V/ \de \vec{X} = \vec{n}$, proving the first statement in Eq. \eqref{toproof}. 

Secondly, we vary the wave front surface 
\begin{align}
S = \int dS = \int \sqrt{h} d^{N-1} \s, \label{actionS}
\end{align}
delivering 
\begin{align}
  \de S =& \int_{\Omega_F} d^{N-1} \s \frac{\de h}{2 \sqrt{h} } 
  	   = \frac{1}{2}\int d^2 \s \sqrt{h} h^{AB} \de h_{AB} \nn \\
=& \int_{\Omega_F} d^{N-1} \s \sqrt{h} h^{AB} \dd_A\vec{X} \cdot  \dd_B \de \vec{X} \nn \\
=& \int_{\Omega_F} d^{N-1} \s \dd_B(\sqrt{h} h^{AB} \dd_A \vec{X}) \cdot \de \vec{X} + B.T.  \nn \\
=& \int_{\Omega_F} dS\  \DD^2 \vec{X} \cdot \de \vec{X}. 
\end{align}
Therefore, $\de S/ \de \vec{X} = \DD^2 \vec{X} = K \vec{n}$ if the boundary term ($B.T.$) vanishes. Indeed, using the divergence theorem with respect to coordinates $\s^A$, we find 
\begin{align}
 B.T. =& \int_{\Omega_F} d^{N-1} \s  \dd_B(\sqrt{h} h^{AB} \dd_A \vec{X} \cdot \de \vec{X} )  \\
=& \int_{\Omega_F} dS  \textrm{div} ( \textrm{grad} \vec{X} \cdot \de \vec{X}) %\nn \\
= \oint_\CC d\ell \ \vec{N} \cdot \de \vec{X} =0.  \nn 
\end{align}
Here, $\vec{N}$ is the domain boundary at the wave front edge $\CC$. Then, by Eq. \eqref{BC}, the boundary term vanishes. Hence we have demonstrated that relation \eqref{cK1} can be written as a gradient system. We conclude that in its motion, a non-linear wave front strives to enlarge its occupied volume, while minimizing its surface area. 

The corollary \eqref{decrease} then states that 
\begin{equation}
 \gamma \dot S - c_0 \dot V = - c_0^2 S - \gamma^2 \int K^2 dS + 2 \gamma c_0 \int K dS \leq 0.
\end{equation}
which is the analogue of the law for vortex filaments by Biktashev et al. \cite{Biktashev:1994}. In particular, for dissipative structures (i.e. when $c_0=0$) which are studied in the context of pattern formation, one has that
\begin{equation}
\dot S = - \gamma \int K^2 dS. 
\end{equation}
Thus, for positive surface tension $\gamma$, dissipative structures evolve towards a minimal surface. 

\textit{Discussion. } We have shown that the geometric law of motion \eqref{cK1} elegantly follows from an action principle that involves only the simplest geometric invariants of a closed surface. In physics, variational principles have allowed to regard the equations of motion in a more abstract manner, and enabled powerful generalizations. Famous examples include the principles of Fermat and D'Alembert, the Hamiltonian and Lagrangian approach to classical mechanics, and their extensions to quantum theory. Other applications of variational principles are the approximation of eigenstates using the Ritz method and imposing constraints in theoretical and numerical problems of fluid dynamics and continuum mechanics.

For non-linear vortices, an extremum principle was found for stationary vortex filaments in \cite{Wellner:2002, tenTusscher:2004}. However, a front with non-vanishing plane-wave speed $c_0$ will never reach a final state unless the wave front itself vanishes. Therefore, it should be of no surprise that the variational principle which we identified here is of the gradient rather than the extremum type. 

At this point, we mention that the terms in the gradient potentials $F$ which we found are highly similar to geometric actions in string theory. Namely, the surface term \eqref{actionS} is identical to the Nambu-Goto string action term, while the volume term \eqref{actionV} has the form of a coupling with an antisymmetric (Kalb-Ramond) field. For propagating dissipation waves, this field happens to be the volume form of the wave front surface~\cite{Zwiebach:2004}.

One may wonder whether a similar variational principle will hold for vortex filaments in extended media. Such filaments can be formed after break-up of a traveling wave front due to the velocity-curvature relation \eqref{cK1}. A scroll wave filament with arc length parameter $s$ is known to obey the geometric law of motion \cite{Keener:1988, Biktashev:1994, Verschelde:2007}:
\begin{align}
 \dot{\vec{X}}(s, t) = \gamma_1 \dd_s^2 \vec{X} + \gamma_2 \dd_s \vec{X} \times \dd_s^2 \vec{X}. \label{eomfil}
\end{align}
If this equation of motion can be written as a gradient system, it needs to satisfy the compatibility condition \eqref{cond}. 
However, for the law \eqref{eomfil}, one computes $ \dot{X}^k/\de X^i = - 2 \gamma_2 \dd_s^3 X^j \eps_{ijm}$, which does not satisfy the criterion unless $\gamma_2=0$. This condition is fulfilled in chemical systems in which all reactants diffuse at the same rate. Thus, the purely dissipative filament dynamics in such systems with no-flux boundary conditions can be written as a gradient system with potential 
\begin{align}
 F = \gamma_1 L,
\end{align}
where $L$ is the total filament length. The corollary \eqref{decrease} recovers the known property that filament length changes monotonically in time \cite{Biktashev:1994}. 

In summary, we have managed to formulate a simple variational principle for the emergent dynamics of propagating non-linear waves in a N-dimensional medium with obstacles. Our result is also a contribution to complexity theory, since we have been able to identify general laws for a wide class of natural systems, without knowing the details of the underlying microscopic processes, and even without knowing the type of partial differential equations that are needed to describe it. 

%\acknowledgements
\textit{Acknowledgements.} H.D. was funded by the FWO-Flanders.

%\bibliographystyle{unsrt}
%\bibliography{../../bib/references_Hans}

\begin{thebibliography}{10}

\bibitem{Zeldovich:1938}
Y.B. Zeldovich and D.A. Frank-Kamenetsky.
\newblock A theory of thermal propagation of flame.
\newblock {\em Acta. Physicochim.}, 274:341--350, 1938.

\bibitem{Nettesheim:1993}
S.~Nettesheim, A.~von Oertzen, H.H. Rotermund, and G.~Ertl.
\newblock Reaction diffusion patterns in the catalytic co‐oxidation on
  pt(110): Front propagation and spiral waves.
\newblock {\em J. Chem. Phys.}, 98:9977--85, 1993.

\bibitem{Zhabotinsky:1973}
A.M. Zhabotinsky and A.N. Zaikin.
\newblock Autowave processes in a distributed chemical system.
\newblock {\em J. Theor. Biol.}, 40:45--61, 1973.

\bibitem{Weinberg:1995}
S.~Weinberg.
\newblock {\em The Quantum Theory of Fields}, volume~2.
\newblock Cambridge University Press, Cambridge, UK, 1995.

\bibitem{Lechleiter:1991}
J.~Lechleiter, S.~Girard, E.~Peraltal, and D.~Clapham.
\newblock Spiral calcium wave propagation and annihilation in {X}enopus
  {L}aevis oocytes.
\newblock {\em Science}, 252:123--126, 1991.

\bibitem{Hodgkin:1952}
A.L. Hodgkin and A.F. Huxley.
\newblock A quantitative description of membrane current and its application to
  conduction and excitation in nerve.
\newblock {\em J. Physiol.}, 117:500--544, 1952.

\bibitem{Klabunde:2004}
R.E. Klabunde.
\newblock {\em Cardiovascular Physiology Concepts}.
\newblock Lippincott Williams \& Wilkins, Baltimore, MD, 2004.

\bibitem{Aslanidi:1997}
O.~V. Aslanidi and O.~A. Mornev.
\newblock Can colliding nerve pulses be reflected?
\newblock {\em Journal of Experimental and Theoretical Physics Letters},
  65(7):579--585, 1997.

\bibitem{Prigogine:1984}
Prigogine I. and Stengers I.
\newblock {\em Order Out of Chaos}.
\newblock Bantam books, New York, 1984.

\bibitem{Hirsch:2004}
Morris~W. Hirsch, Stephen Smale, Robert~L. Devaney, and Morris~W. Hirsch.
\newblock {\em Differential equations, dynamical systems, and an introduction
  to chaos}.
\newblock Academic Press, San Diego, CA, 2nd ed edition, 2004.

\bibitem{Mornev:1998}
O.A. Mornev.
\newblock Modification of the biot method on the basis of the principle of
  minimum dissipation (with an application to the problem of nonlinear
  concentration waves in an autocatalytic medium).
\newblock {\em Russ J Phys Chem}, 72:112--118, 1998.

\bibitem{Frankel:1997}
T.~Frankel.
\newblock {\em The geometry of physics}.
\newblock Cambridge University Press, Cambridge, 1997.

\bibitem{Mornev:1992}
O.A. Mornev, A.V. Panfilov, and R.R. Aliev.
\newblock {Fitzhugh-Nagumo equations are a gradient system}.
\newblock {\em {Biofizika}}, {37}({1}):{123--125}, {1992}.

\bibitem{Yanagida:2002}
E.~Yanagida.
\newblock Mini-maximizers for reaction-diffusion systems with skew-gradient
  structure.
\newblock {\em J. Differential Equations}, 179:311--335, 2002.

\bibitem{Keener:1986}
J.P. Keener.
\newblock A geometrical theory for spiral waves in excitable media.
\newblock {\em Siam J Appl Math}, 46:1039--1056, 1986.

\bibitem{Dierckx:2011}
H.~Dierckx, O.~Bernus, and H.~Verschelde.
\newblock Accurate eikonal-curvature relation for wave fronts in locally
  anisotropic reaction-diffusion systems.
\newblock {\em Phys Rev Lett}, 107:108101, 2011.

\bibitem{Panfilov:1995b}
A.V. Panfilov and J.P. Keener.
\newblock Dynamics of dissipative structures in reaction-diffusion equations.
\newblock {\em SIAM J. Appl. Math.}, 55:205--219, 1995.

\bibitem{Verschelde:2007}
H.~Verschelde, H.~Dierckx, and O.~Bernus.
\newblock Covariant stringlike dynamics of scroll wave filaments in anisotropic
  cardiac tissue.
\newblock {\em Phys. Rev. Lett.}, 99:168104, 2007.

\bibitem{Young:2010}
R.J. Young and A.V. Panfilov.
\newblock Anisotropy of wave propagation in the heart can be modeled by a
  riemannian electrophysiological metric.
\newblock {\em Proc Natl Acad Sci USA}, 107:15063--8, 2010.

\bibitem{Biktashev:1994}
V.N. Biktashev, A.V. Holden, and H.~Zhang.
\newblock Tension of organizing filaments of scroll waves.
\newblock {\em Phil. Trans. R. Soc. Lond. A}, 347:611--630, 1994.

\bibitem{Wellner:2002}
M.~Wellner, O.M. Berenfeld, J.~Jalife, and A.M. Pertsov.
\newblock Minimal principle for rotor filaments.
\newblock {\em P Natl Acad Sci USA}, 99:8015--8018, 2002.

\bibitem{tenTusscher:2004}
K.H.W. ten Tusscher and A.V. Panfilov.
\newblock Eikonal formulation of the minimal principle for scroll wave
  filaments.
\newblock {\em Phys. Rev. Lett.}, 93:108106, 2004.

\bibitem{Zwiebach:2004}
B.~Zwiebach.
\newblock {\em A first course in string theory}.
\newblock Cambridge University Press, Cambridge, 2004.

\bibitem{Keener:1988}
J.P. Keener.
\newblock The dynamics of three-dimensional scroll waves in excitable media.
\newblock {\em Physica D}, 31:269--276, 1988.

\end{thebibliography}

\end{document}